
\input phyzzx
\scrollmode

\def\ch{{\cal H}} \def\r{\rangle} \def\l{\langle}
\def\b{\beta} \def\g{\gamma} \def\p{\partial}
\def\bc{\bar c} \def\bL{\bar L} \def\rn{R^{(N)}}
\def\fn{f^{(N)}} \def\c{\circ} \def\eib{(\eta_i|B)}
\def\hV{\hat V} \def\hP{\hat\Phi} \def\co{{\cal O}}
\def\hPs{\hat\Psi} \def\al{{(\alpha)}} \def\cb{{\cal B}}

\overfullrule=0pt
{}~\hfill\vbox{\hbox{TIFR-TH-92/14}\hbox{February, 1992}}

\title{PICTURE CHANGING OPERATORS IN CLOSED FERMIONIC STRING FIELD THEORY}

\author{R. Saroja and Ashoke Sen\foot{e-mail addresses:
RSAROJA@TIFRVAX.BITNET, SEN@TIFRVAX.BITNET}}

\address{Tata Institute of Fundamental Research, Homi Bhabha Road, Bombay
400005, India}

\abstract

We discuss appropriate arrangement of picture changing operators required
to construct gauge invariant interaction vertices involving
Neveu-Schwarz states in heterotic and closed superstring field theory.
The operators required for this purpose are shown to satisfy a set of
descent equations.

\let\refmark=\NPrefmark 
\def\define#1#2\par{\def#1{\Ref#1{#2}\edef#1{\noexpand\refmark{#1}}}}
\def\con#1#2\noc{\let\?=\Ref\let\<=\refmark\let\Ref=\REFS
         \let\refmark=\undefined#1\let\Ref=\REFSCON#2
         \let\Ref=\?\let\refmark=\<\refsend}

\endpage

\define\SAPOL
M.~Saadi and B.~Zwiebach, Ann.~Phys. {\bf 192} (1989) 213.

\define\KUPOL
T.~Kugo, H.~Kunitomo, and K.~Suehiro, Phys.~Lett. {\bf 226B} (1989) 48.

\define\WITTEN
E. Witten, Nucl. Phys. {\bf B268} (1986) 253.

\define\LPP
A. LeClair, M. Peskin and C. Preitschopf, Nucl. Phys. {\bf B317} (1989)
411,464; S. Samuel, Nucl. Phys. {\bf B308} (1988) 317;
A.~Sen, Nucl.~Phys. {\bf B335} (1990) 435.

\define\BIG
N. Berkovits, M. Hatsuda, and W. Siegel, preprint ITP-SB-91-36.

\define\THORN
C. Preitschopf, C. Thorn and S. Yost, Nucl. Phys. {\bf B337} (1990) 363.

\define\MATV
I. Arafeva, P. Medvedev and A. Zubarev, Nucl. Phys. {\bf B341} (1990) 364;
Phys. Lett. {\bf B240} (1990) 356.

\define\VV
E. Verlinde and H. Verlinde, Phys. Lett. {\bf B192} (1987) 95.

\define\WENDT
C. Wendt, Nucl. Phys. {\bf B314} (1989) 209.

\define\SAMUEL
O. Lechtenfeld and S. Samuel, Nucl. Phys. {\bf B310} (1988) 254; Phys.
Lett. {\bf B213} (1988) 431; R. Bluhm
and S. Samuel, Nucl. Phys. {\bf B338} (1990) 38.

\define\AMS
J. Atick, G. Moore and A. Sen, Nucl. Phys. {\bf B308} (1988) 1.

\define\KAKU
M. Kaku and J. Lykken, Phys. Rev. {\bf D38} (1988) 3067;
M. Kaku, preprints CCNY-HEP-89-6, Osaka-OU-HET 121.

\define\GAUGEINV
T.~Kugo and K.~Suehiro, Nucl.~Phys. {\bf B337} (1990) 434.

\define\BACKGND
A.~Sen, Nucl.~Phys. {\bf B345} (1990) 551.

\define\ZWIEBACH
B.~Zwiebach, Mod.~Phys.~Lett. {\bf A5} (1990) 2753;
Phys. Lett. {\bf B241} (1990) 343; Comm. Math. Phys. {\bf 136} (1991) 83.

\define\MARTINEC
E.~Martinec, Nucl.~Phys. {\bf B281} (1987) 157.

\define\FMS
D. Friedan, E. Martinec and S. Shenker, Phys. Lett. {\bf B160} (1985) 55;
Nucl. Phys. {\bf B271} (1986) 93.


A  closed bosonic string field theory based on non-polynomial
interaction has been constructed
recently\con\SAPOL\KUPOL\GAUGEINV\KAKU\noc.
In this paper we shall discuss the construction of gauge invariant field
theory for heterotic and type II superstring theories based on the same
principles.
However, our analysis will be confined only to the Neveu-Schwarz sector of
the theory; the construction of a similar field theory including Ramond
sector states involves extra complications due to problems involving zero
modes of various fields.
Although the field theory for fermionic strings is not complete without
Ramond sector, we can get some insight into the theory just by looking at
the field theory involving the Neveu-Schwarz states.
In particular, since for the heterotic string theory, all the bosonic
fields come from the Neveu-Schwarz sector, the field theory involving
Neveu-Schwarz states can be used to study the space of classical
solutions, as well as the geometry of the configuration space of the
theory.


We shall, for convenience, restrict our discussion to heterotic string
theory only, the analysis for superstring theory proceeds in a similar
manner.
Let $\ch$ denote the Hilbert space of states in the Neveu-Schwarz sector
in the $-1$ picture.
These are the states created by the matter operators, and the
ghost oscillators $b_n$, $c_n$, $\b_n$, $\g_n$ acting on the state
$e^{-\phi(0)}|0\r \equiv |\Omega\r$,
where $\phi$ is the bosonized ghost, related to $\b$, $\g$ through the
relations $\b= e^{-\phi}\p\xi$, $\g =e^{\phi}\eta$.
Here $\eta$ and $\xi$ are fermionic fields of dimensions $1$  and 0
respectively.
$|0\r$ denotes the SL(2,C) invariant vacuum in the combined matter ghost
theory.
As in the case of bosonic string theory, a general off-shell string state
$|\Psi\r$ is taken to be a GSO projected state in $\ch$ annihilated by
$c_0^-\equiv
(c_0-\bc_0)/\sqrt 2$ and $L_0^-\equiv (L_0-\bL_0)/\sqrt 2$, and created
from $|\Omega\r$ by an operator of total ghost number 3.
Following the construction of bosonic string field theory, we shall look
for an action of the heterotic string field theory of the form:
$$
S(\Psi) = {1\over 2} \l\Psi|Q_B b_0^-|\Psi\r +\sum_{N=3}^\infty
{g^{N-2}\over N!} \{\Psi^N\}
\eqn\ethree
$$
where $\{ A_1\ldots A_N\}$ and $[A_1\ldots A_N]$ are multilinear maps from
$N$-fold tensor product of $\ch$ to $C$ (the space of complex numbers) and
$\ch$ respectively, satisfying
relations identical to the corresponding relations in bosonic string field
theory:
$$
\{ A_1\ldots A_N\} \equiv (-1)^{n_1+1}\l A_1|[A_2\ldots A_N]\r
\eqn\efour
$$
$$
\{A_1\ldots A_N\} =(-1)^{(n_i+1)(n_{i+1}+1)}\{A_1 \ldots A_{i-1} A_{i+1}
A_i A_{i+2} \ldots A_N\}
\eqn\efive
$$
$$
\{A_1 \ldots (L_0^- A_i) \ldots A_N\} = 0 = \{A_1\ldots (b_0^- A_i)\ldots
A_N\}
\eqn\eseven
$$
$$\eqalign{
(-1)^{n_1}&\{(Q_B A_1)A_2\ldots A_N\} = \sum_{i=2}^N
(-1)^{\sum_{j=2}^{i-1} (n_j+1)} \{A_1\ldots (Q_B A_i)\ldots A_N\}\cr
&-\sum_{m,n\ge 3\atop m+n=N+2}\sum_{\{j_k\}, \{i_l\}}
(-1)^{\sigma(\{i_l\}, \{j_k\})}\{ A_1 A_{j_1}\ldots A_{j_{m-2}}
c_0^-[A_{i_1}\ldots A_{i_{n-1}}]\}\cr
}
\eqn\etwenty
$$\def\eeight{\etwenty}
Here $n_i$ denotes the ghost number of the state $|A_i\r$.
The sum over $\{i_l\}$, $\{j_k\}$ in eq.\eeight\ runs over all possible
divisions of the set of integers $2,\ldots N$ into the sets $\{i_l\}$ and
$\{j_k\}$.
$(-1)^{\sigma(\{i_l\}, \{j_k\})}$ is a factor of $\pm 1$ which is computed
as follows.
Starting from the ordering $Q_B, A_2,\ldots A_N$, bring them into the
order $A_{j_1},\ldots A_{j_{m-2}}, Q_B, A_{i_1},\ldots A_{i_{n-1}}$, using
the rule that $Q_B$ is anticommuting, and $A_i$ is commuting
(anti-commuting) if $n_i$ is odd (even).
The sign picked up during this rearrangement is $(-1)^{\sigma(\{i_l\},
\{j_k\})}$.
Using eqs.\efour-\eeight\ and the nilpotence of the BRST charge $Q_B$, one
can show that $S(\Psi)$ given in eq.\ethree\ is invariant under a gauge
transformation of the form:
$$
b_0^- \delta|\Psi\r = Q_B b_0^- |\Lambda\r +\sum_{N=3}^\infty
{g^{N-2}\over (N-2)!} [\Psi^{N-2}\Lambda]
\eqn\enine
$$
where $|\Lambda\r$ is an arbitrary GSO projected state in $\ch$ created
from $|\Omega\r$
by an operator of ghost number 2, and annihilated by $c_0^-$ and $L_0^-$.

Thus, in order to construct a gauge invariant action, we need to construct
multilinear maps $\{A_1\ldots A_N\}$ from $\ch^N$ to $C$, satisfying
eqs.\efour-\eeight.
We look for an expression similar to the one in bosonic string
theory\GAUGEINV\BACKGND:
$$
\{A_1\ldots A_N\} =-\int_{\rn} d^{2N-6}\tau \l \fn_1\c (b_0^- P A_1)(0)
\ldots \fn_N\c (b_0^- PA_N)(0) K_N\r
\eqn\eten
$$
Here $P$ is the projection operator $\delta_{L_0\bL_0}$.
$\fn_i$ is a
conformal map that maps the unit circle into the $i$th external string of
the $N$-string diagram associated with the $N$-string vertex.
$\fn_i\c (b_0^- PA_i)$ denotes the conformal transform of the field
$(b_0^-PA_i)$ under the map $\fn_i$.
$\tau^i$ are the modular parameters characterizing the $N$-string diagram.
The region of integration $\rn$, and the $N$-string diagram is chosen in
such a way that at the boundary $\p\rn$ of $\rn$, the $N$-string diagram
is identical to an $m$ string diagram and an $n(=N+2-m)$ string diagram
glued by a tube of zero length and twist $\theta$.
One particular example of such $N$-string diagrams is provided by
polyhedra with $N$-faces, the perimeter of each face being equal to
$2\pi$.
In this case $\tau_i$ correspond to independent parameters labelling the
lengths of each side of the polyhedron.
The region of integration $\rn$ over $\tau$ is such that it includes all
such polyhedra with the restriction that the length of any closed curve on
the polyhedron constructed out of the edges should be greater than or
equal to $2\pi$.
(Such polyhedra were called regular polyhedra in ref.\KUPOL).

Finally we turn to the description of the operator $K_N$ appearing in
eq.\eten.
For bosonic string field theory $K_N$ was given by $\prod_{i=1}^{2N-6}\eib$,
where
$$
\eib = \int d^2 z (\eta_{i\bar z}^{~~z} b(z) +\bar\eta_{iz}^{~~\bar z}
\bar b(\bar z))
\eqn\eeleven
$$
$\eta_{i\bar z}^{~~z}$, $\bar\eta_{iz}^{~~\bar z}$ are the beltrami
differentials, which tell us how the components $g^{zz}$, $g^{\bar z\bar
z}$ of the metric, induced on the sphere by the $N$-string diagram,
changes as we change $\tau^i$\GAUGEINV.
In the case of fermionic string theories, such a choice of $K_N$ will give
vanishing answer for $\{A_1\ldots A_N\}$ due to ghost number
non-conservation.
This is remedied by introducing appropriate factors of picture changing
operators\con\FMS\VV\BIG\noc in the definition of $K_N$\WITTEN.
We shall make the following choice of $K_N$:
$$
K_N=\sum_{r=0}^{2N-6} C^{(r)}_N\wedge \Phi_N^{(2N-6-r)}
\eqn\etwelve
$$
where $C^{(r)}_N$ is an $r$-form (in the $2N-6$ dimensional moduli space
spanned by $\tau^i$) operator, given by,
$$
\big(C^{(r)}_N\big)_{i_1\ldots i_r} =\prod_{k=1}^r (\eta_{i_k}|B)
\eqn\ethirteen
$$
$\Phi_N^{(r)}$, on the other hand, is an $r$ form operator satisfying the
`descent' equation:
$$
d^{(E)}\Phi_N^{(r)} =[Q_B,\Phi_N^{(r+1)}\}
\eqn\eseventeen
$$
where $d^{(E)}$ denotes derivative with respect to $\tau^i$, acting on the
explicit $\tau$ dependent factors in the expressions for $\Phi_N^{(r)}$,
and $[~\}$ denotes a commutator (anti-commutator) for odd (even) $r$ in
eq.\eseventeen.
$\Phi_N^{(0)}$ takes the form:
$$
\Phi_N^{(0)} = \sum_\alpha A^{(\alpha)}(\tau^1,\ldots \tau^{2N-6})
X(w_1^{(\alpha)}(\tau)) \ldots X(w_{N-2}^{(\alpha)}(\tau))
\eqn\efourteen
$$
where $X(z)=\{Q_B,\xi(z)\}$ is the picture changing operator\FMS.
The sum over $\alpha$ in eq.\efourteen\ runs over a finite set of values.
For each value of $\alpha$ we have a set of points $w_1^{(\alpha)}(\tau),
\ldots w_{N-2}^{(\alpha)}(\tau)$ on the $N$-string diagram, and an weight
factor $A^{(\alpha)}(\tau)$, satisfying the normalization condition
$\sum_\alpha A^{(\alpha)}(\tau) =1$.
$\Phi_N^{(r)}$'s, on the other hand, are constructed as linear
combinations of products of $X$'s, $\p\xi$'s and
$(\xi(P)-\xi(Q))$'s.\foot{ Appearance of extra factors proportional to
$\p\xi$ when the locations of the picture changing operators are moduli
dependent, was discussed in ref.\AMS.}

Besides the descent equation \eseventeen, $\Phi_N^{(r)}$ are also required
to satisfy the following boundary conditions.
As has been indicated before, the boundary $\p\rn$ of $\rn$ consists of
several pieces; each piece corresponds to gluing two vertices with less
number of external states ($m$ and $n=N+2-m$, say) along one
of the external strings from each vertex with a certain twist $\theta$.
We demand that on such a component of $\p\rn$,
$$
\Phi_N^{(r)}\bigg|_{\p\rn} =\sum_{s=0}^r \Phi_m^{(r-s)}\wedge \Phi_n^{(s)}
\eqn\eeighteen
$$
Here $\Phi_N^{(r)}\bigg|_{\p\rn}$ denotes the component of $\Phi_N^{(r)}$
tangential to $\p\rn$.
Finally, $\Phi_N^{(r)}$ should be invariant under permutation of the
external strings.
Besides these restrictions, the choice of $\Phi_N^{(r)}$ (i.e. the
quantities $A^{(\alpha)}$, $w_i^{(\alpha)}$, and the corresponding
quantities appearing in the expressions for $\Phi_N^{(r)}$ for $r\ge 1$)
is completely arbitrary.\foot{ Although we shall try to choose
$\Phi_N^{(r)}$ in such a way that they are continuous inside $\rn$,
this is not a necessary constraint, since, as can be
seen from eq.\eseventeen, a discontinuity of $\Phi_N^{(r)}$ inside $\rn$
may be compensated by a $\delta$-function singularity in
$\Phi_N^{(r+1)}$.}

We shall now show that the quantities $\{A_1\ldots A_N\}$ constructed this
way are non-zero in general, and satisfy eqs.\efive-\eeight\ with
$[A_1\ldots A_{N-1}]$ defined through eq.\efour.
To see that $\{A_1\ldots A_N\}$ is non-zero in general, we only need to
note that the contribution to the right hand side of eq.\etwelve\ from the
$r=2N-6$ term is of the form
$$
\prod_{i=0}^{2N-6} \eib\sum_\alpha A^{(\alpha)}(\tau) X(w_1^{(\alpha)})
\ldots X(w_{N-2}^{(\alpha)})
\eqn\eextraa
$$
Besides providing the appropriate factors of $\eib$ as in the case of
bosonic string theory, we now also have the correct number of picture
changing operators.
Thus, if each $|A_i\r$ is created by a ghost number 3
operator acting on $|\Omega\r$, at least the contribution from the
$r=2N-6$ term in eq.\etwelve\ to $\{A_1\ldots A_N\}$ will be non-zero.
As we shall now see, the other terms are necessary for $\{A_1\ldots A_N\}$
to satisfy eq.\eeight.

Verification of eq.\efour-\eseven\ with the definition of $\{A_1\ldots
A_N\}$ given in eq.\eten\ is straightforward, so we turn to the
verification of eq.\eeight.
Using eq.\eten, both sides of eq.\etwenty\ may be expressed as integrals
(over $\tau^i$) of appropriate correlation functions in the conformal
field theory.
In the left hand side of eq.\etwenty, we may express $(Q_BA_1)$ as the
contour integral of the BRST current around the location of the operator
$b_0^-A_1$.
We may now deform the BRST contour and shrink it to a point, in the
process picking up residues from the locations of various other operators.
The residues from the locations of the operators $b_0^-A_i$ ($i\ge 2$)
give rise to the first set of terms on the right hand side of
eq.\etwenty.
We are now left with terms proportional to $[Q_B,K_N]$.
Residues from the terms proportional to $b(z)$, $\bar b(\bar z)$ in $\eib$
generates a factor of $\int d^2 z(\eta_{i\bar z}^{~~ z} T(z)
+\bar\eta_{iz}^{~~\bar z}\bar T(\bar z))\equiv T^{(1)}_i$.
Thus we may write,
$$
[Q_B, C_N^{(r)}\} =T^{(1)}\wedge C_N^{(r-1)}
\eqn\etwentyone
$$
where $C_N^{(r)}$ has been defined in eq.\ethirteen.
Since insertion of a $T^{(1)}_i$ inside a correlation function generates
the $\tau^i$ derivative of the correlation function, with the derivative
acting on the implicit $\tau^i$ dependence of the correlation function due
to the dependence of the string diagram on $\tau$, we see that the terms
proportional to $[Q_B,C^{(r)}_N\}$ in $[Q_B,K_N]$ give rise to a term of
the form:
$$
\int_{\rn} d^{2N-6}\tau d^{(I)} \cb
\eqn\etwentytwo
$$
where,
$$
\cb=(-1)^{n_1}\l \prod_{i=1}^N (\fn_i\c b_0^- A_i(0)) \sum_{r=1}^{2N-6}
C_N^{(r-1)} \wedge \Phi_N^{(2N-6-r)}\r
\eqn\etwentythree
$$
and $d^{(I)}$ denotes the $\tau$ derivative acting on the implicit $\tau$
dependence of the correlation function but not on the explicit $\tau$
dependence appearing in $\Phi_N^{(r)}$.
On the other hand, terms proportional to $[Q_B,\Phi_N^{(r)}\}$ in
$[Q_B,K_N]$ may be analyzed using the descent equation \eseventeen, and
gives an answer similar to eq.\etwentytwo\ with $d^{(I)}$ replaced by
$d^{(E)}$.
Thus if $d\equiv d^{(I)}+d^{(E)}$ denotes the total $\tau$ derivative, the
contribution from the term proportional to $[Q_B,K_N\}$ that appears in
the analysis of the left hand side of eq.\etwenty\ may be written as,
$$
\int_{\rn} d^{2N-6}\tau d\cb =\int_{\p\rn} d^{2N-7}\tau \cb
\eqn\etwentyfour
$$

Let us now consider a specific component of the boundary $\p\rn$ of $\rn$
that corresponds to gluing of two lower order string diagrams of $m$ and
$n=N+2-m$ external states,
with a twist $\theta$.
Let $\tau_{(1)}$ and $\tau_{(2)}$ be the modular parameters describing
these lower order string diagrams.
Then,
$$
d^{2N-7}\tau\bigg|_{\p\rn} = d^{2m-6}\tau_{(1)} d^{2n-6}\tau_{(2)} d\theta
\eqn\etwentyfive
$$
$$
\Big(C_N^{(r)}\Big)_{\theta a_1\ldots a_{r-1}}\Big|_{\p\rn} =
(\eta_\theta|B)\Big(\sum_{s=0}^{r-1}
C_m^{(s)} \wedge C_n^{(r-s-1)}\Big)_{a_1\ldots a_{r-1}}
\eqn\etwentysix
$$
Using the boundary conditions given in eqs.\eeighteen\ and \etwentysix\ it
is easy to see that,
$$\eqalign{
&\Big(\sum_{r=1}^{2N-6} C_N^{(r-1)}\wedge \Phi_N^{(2N-6-r)}\Big)_{
\theta\tau_{(1)}^1 \ldots \tau_{(1)}^{2m-6}\tau_{(2)}^1\ldots
\tau_{(2)}^{2n-6} }\Big|_{\p\rn}\cr
=& (\eta_\theta|B) \Big(\sum_{r=0}^{2m-6} C_m^{(r)}\wedge
\Phi_m^{(2m-6-r)} \Big)_{\tau_{(1)}^1\ldots \tau_{(1)}^{2m-6} }
\Big(\sum_{s=0}^{2n-6} C_n^{(s)}\wedge
\Phi_n^{(2n-6-s)} \Big)_{\tau_{(2)}^1\ldots \tau_{(2)}^{2n-6}} \cr
}
\eqn\etwentyseven
$$
Standard manipulations identical to the one for the bosonic string theory
can now be used to show that the contribution to the right hand side of
eq.\etwentyfour\ is identical to the second set of terms to the right hand
side of eq.\etwenty.
This completes the proof of eq.\etwenty.


We shall now indicate the
basic steps involved in the calculation of Feynman amplitudes in this
field theory.
The external states are taken to be physical states of the form
$b_0^-|A_i\r = c\bar c V_i(0)|\Omega\r\equiv c\bar c\hV_i(0)|0\r$
where $V_i$ is a dimension $(1/2, 1)$ superconformal primary field in the
matter sector.
Since the contribution from various Feynman diagrams can be brought into
the form of the contribution from elementary $N$-point vertex given in
eq.\eten\ using standard techniques\LPP, we shall analyze the contribution
to $A(1,\ldots N)$ from the elementary $N$-point vertex only.
In analyzing this contribution we shall use the familiar expression for
$\eta_{i\bar z}^{~~z}$, $\bar\eta_{iz}^{~~\bar z}$ in terms of quasi
conformal deformations\MARTINEC\ $v^z$, $\bar v^{\bar z}$, and write,
$$
(\eta_i|B)=\ointop_C (dz {\delta v^z\over\delta\tau^i} b(z) +d\bar
z{\delta \bar v^{\bar z}\over\delta\tau^i} \bar b(\bar z) )
\eqn\etwentyseveng
$$
where the contour $C$ encloses all the points $z_i$, as well as the
locations of all the operators appearing in $\hP_N^{(r)}$.
$\delta v^z$ ($\delta\bar v^{\bar z}$) is analytic (anti-analytic) inside
the contour $C$ but not outside $C$.
We can now deform $C$ and shrink it to a point, picking up residues from
the locations of various operators in this process.
In the case of bosonic string theory, the only possible residues are
picked up from the locations of the vertex operators $b_0^- A_i=c\bar c
\hV_i$.
This removes the $c\bar c$ factors from $N-3$ of the vertex operators and
generates appropriate measure factors that converts the integration over
$\tau^i$ to integration over the locations of the $(N-3)$ vertices.
In the present case, however, the integration contour can pick up residues
from the locations of the picture changing operators inside $\Phi_N^{(r)}$
also.
For any $r$ form operator $\co$, let us define an $r+1$ form operator
$\delta\co$ as,
$$
(\delta\co)_{i_1\ldots i_{r+1}} = A\Big(\ointop_C \Big(
dz {\delta v^z\over\delta\tau^{i_1}} b(z) +d\bar
z{\delta \bar v^{\bar z}\over\delta\tau^{i_1}} \bar b(\bar z)\Big)
\co_{i_2\ldots i_{r+1}}\Big)
\eqn\etwentysevenh
$$
where $A$ denotes antisymmetrization in the indices $i_1, \ldots i_{r+1}$,
and $C$ denotes a contour enclosing the locations of all the operators in
$\co$.
Using eqs.\eten-\ethirteen\ the contribution to $A(1,\ldots N)$ from the
elementary $N$-point vertex may be written as,
$$
g^{N-2}\int_{\rn}  d^{2N-6}\tau \sum_{r=0}^{2N-6}\sum_{s=0}^r {1\over s!
(r-s)!}
\Big\l\delta^s\Big(\prod_{k=1}^N c\bar c\hV_k(z_k,\bar z_k) \Big)\wedge
\delta^{r-s}\Phi_N^{(2N-6-r)} \Big\r
\eqn\etwentyseveni
$$
where $\delta^s$ denotes $s$ successive operations of $\delta$ and
$z_i=f^N_i(0)$.

So far our description has been independent of the choice of coordinates
of the moduli space.
We can now simplify our analysis by choosing the moduli parameters
$\tau^i$ to be identical to the coordinates $x_i$ ($1\le i\le 2N-6$)
defined as follows:
$$
x_i = z_{i+1\over 2}~~~{\rm for}~i~{\rm odd}; ~~~~~
x_i=\bar z_{i\over 2}~~~{\rm for}~i~{\rm even}
\eqn\etwentysevenc
$$
In that case, for odd $i$,
$$\eqalign{
{\delta v^z\over\delta\tau^i} =& 1~~{\rm at}~z=z_{i+1\over 2};~~~~
{\delta v^z\over\delta\tau^i} = 0~~{\rm at}~z=z_j~{for}~j\ne {i+1\over
2}\cr
{\delta\bar v^{\bar z}\over\delta\tau^i} =& 0~~{\rm at}~z=z_j~{\rm
for~all}~ j\cr
}
\eqn\etwentysevenj
$$
and, for even $i$,
$$\eqalign{
{\delta v^z\over\delta\tau^i} =& 0~~{\rm at}~z=z_j~{\rm
for~all}~ j\cr
{\delta\bar v^{\bar z}\over\delta\tau^i} =& 1~~{\rm at}~z=z_{i\over
2};~~~~
{\delta\bar v^{\bar z}\over\delta\tau^i} =0~~{\rm at}~z=z_j~{\rm
for}~j\ne{i\over 2}\cr
}
\eqn\etwentysevenk
$$
Let us now define,
$$
\hP_N^{(s)} =\sum_{r=0}^s {1\over r!} \delta^r\Phi_N^{(s-r)}
\eqn\etwentysevenm
$$
and,
$$
L_N^{(s)} = {1\over s!}\delta^s\Big(\prod_{k=1}^N c\bar c\hV_k)
\eqn\etwentysevenl
$$
so that,
$$
\Big(L_N^{(2N-6-r)}\Big)_{i_1\ldots i_{2N-6-r}} = {\delta\over\delta
C_{i_1}} \ldots {\delta\over\delta C_{i_{2N-6-r}}} \prod_{i=1}^N
c\bar c\hV_i(z_i,\bar z_i)
\eqn\etwentysevend
$$
where,
$$
C_j = c(z_{j+1\over 2})~~~{\rm for}~j~{\rm odd};~~~~
C_j= \bar c(\bar z_{j\over 2})~~~{\rm for}~j~{\rm even}
\eqn\etwentysevene
$$
Then eq.\etwentyseveni\ takes the form:
$$
A(1,\ldots N) = g^{N-2}\int d^{2N-6} x_i\sum_{r=0}^{2N-6} \l
L_N^{(r)}\wedge \hP_N^{(2N-6-r)} \r
\eqn\etwentysevenb
$$
By techniques identical to those used in the case of bosonic string field
theory one can show that the integral over $x_i$ runs over the full moduli
space when we add the contribution from all the Feynman diagrams.

Using the relations $\{Q_B, b(z)\}=T(z)$, $\{Q_B,\bar
b(\bar z)\}=\bar T(\bar z)$, the fact that the insertion of $\int
d^2z(\eta_{i\bar z}^{~~z}T(z)+\bar\eta_{iz}^{~~\bar z}\bar T(\bar z))$ in
a correlation function generates the (implicit) derivative of the
correlation function with respect to $\tau^i$, and the descent equation
\eseventeen\ we get,
$$
[Q_B, \delta^r\Phi_N^{(l)}\} = r d^{(I)}(\delta^{r-1}\Phi_N^{(l)})
+d^{(E)}(\delta^r\Phi_N^{(l-1)})
\eqn\etwentysevenp
$$
Using eq.\etwentysevenp\ one can easily verify that
$\hP_N^{(r)}$ defined in eq.\etwentysevenm\ satisfies the descent
equation:
$$
d\hP_N^{(r)} =[Q_B,\hP_N^{(r+1)}\}
\eqn\etwentysevenf
$$
where $d$ denotes total derivative with respect to the $x_i$'s.

We now turn to the question of comparing the amplitude given in
eq.\etwentysevenb\ with the one calculated from the first quantized
formalism.
In order to do so we shall first prove the following

\noindent {\it Lemma: If we have two sets of operators $\hP_N^{(r)}$ and
$\hPs_N^{(r)}$, both satisfying the descent equations given in
eq.\etwentysevenf, and, if,
$$
\hP_N^{(0)} -\hPs_N^{(0)} =\{Q_B, \chi_N^{(0)}\}
\eqn\etwentysevenq
$$
for some $ \chi_N^{(0)}$, then,
$$
\int d^{2N-6} x_i \sum_{r=0}^{2N-6} \l L_N^{(r)}\wedge (\hP^{(2N-6-r)}_N
-\hPs^{(2N-6-r)}_N)\r =0
\eqn\etwentysevenr
$$
}

The lemma is proved in the following way.
{}From eq.\etwentysevenq\ and the descent equation for $\hP_N^{(r)}$,
$\hPs_N^{(r)}$, we get,
$$
\{Q_B,\hP_N^{(1)} -\hPs_N^{(1)}\} =\{Q_B, d \chi_N^{(0)}\}
\eqn\etwentysevens
$$
This gives,
$$
\hP_N^{(1)} -\hPs_N^{(1)} = d\chi_N^{(0)} +[Q_B, \chi_N^{(1)}]
\eqn\etwentysevent
$$
for some $\chi_N^{(1)}$.
Repeating this, we get the general equation,
$$
\hP_N^{(r)} -\hPs_N^{(r)} = d\chi_N^{(r-1)} +[Q_B,\chi_N^{(r)}\}
\eqn\etwentysevenu
$$
Finally, using the definition of $L_N^{(r)}$ and the commutation relations,
$$\eqalign{
[Q_B,\hV_i(z_i,\bar z_i)] =& {\p\over\p z_i}\big(c\hV_i(z_i,\bar z_i)\big)
+{\p\over\p\bar z_i}\big(\bar c\hV_i(z_i,\bar z_i)\big)\cr
\{Q_B,c\hV_i(z_i,\bar z_i)\} =&{\p\over \p\bar z_i}(\bar c c\hV_i),~~
\{ Q_B,\bar c\hV_i\}={\p\over\p z_i}(c\bar c\hV_i), ~~ [Q_B,c\bar
c\hV_i]=0 \cr
}
\eqn\etwentysevenv
$$
we get,
$$
[Q_B, L_n^{(r)}\} = d L_N^{(r-1)}
\eqn\etwentysevenw
$$
Using eqs.\etwentysevenu, \etwentysevenw, and the fact that the
expectation value of a BRST exact operator vanishes, we can bring the
left hand side of eq.\etwentysevenr\ into the form:
$$
\int d^{2N-6} x_i d\Big(\sum_{r=0}^{2N-7} L_N^{(r)}\wedge
\chi_N^{(2N-7-r)} (-1)^{r}\Big)
\eqn\etwentysevenx
$$
which vanishes after integration over $x_i$.
This completes the proof of the lemma.

Let us now choose,
$$
\hPs_N^{(0)} = X(z_3) X(z_4)\ldots X(z_N)
\eqn\etwentyseveny
$$
Using the fact that $\hP_N^{(0)}$ has the form $\sum_\alpha\tilde
A^{(\alpha)}(\{x_i\}) X(w_1^{(\alpha)})\ldots X(w_{N-2}^{(\alpha)})$ with
$\sum_\alpha \tilde A^{(\alpha)}=1$, we see that
$\hP_N^{(0)}-\hPs_N^{(0)}$ is BRST exact.
Thus by the above lemma, we can replace $\hP_N^{(r)}$ by $\hPs_N^{(r)}$ in
the expression for $A(1,\ldots N)$ given in eq.\etwentysevenb.
A set of $\hPs_N^{(r)}$ satisfying the descent equation are given by,
$$\eqalign{
\big(\hPs_N^{(r)}\big)_{i_1\ldots i_r} =&\prod_{k=1}^r\p\xi(z_{i_k+1\over
2}) \prod_{j=3\atop j\ne {i_l+1\over 2}}^N X(z_j)~~{\rm for}~i_k~{\rm
odd~and} ~3\le {i_k+1\over 2}\le N-3~{\rm for~all}~k\cr
=& 0~~{\rm otherwise}\cr
}
\eqn\etwentysevenz
$$
Using this we get the following expression for
$A(1,\ldots N)$:
$$\eqalign{
A(1,\ldots N) =& g^{N-2}\int \prod_{i=1}^{N-3} d^2
z_i\prod_{i=1}^2\hV_i(z_i,\bar z_i)\prod_{i=N-2}^N(c\bar c\hV_i(z_i, \bar
z_i) X(z_i))\cr
& \prod_{i=3}^{N-3}\Big(\hV_i(z_i,\bar z_i)(X(z_i)-c(z_i)\p\xi(z_i))\Big)\cr
}
\eqn\etwentysevenza
$$
where $d^2 z_i\equiv d\bar z_i\wedge dz_i$.
In the above equation, $\hV_i(z_i,\bar z_i)$ corresponds to integrated
vertex operator in the $-1$ picture\FMS.
Also,
$$
\hV_i(z_i,\bar z_i)(X(z_i)-c(z_i)\p\xi(z_i)) =\{Q_B,\hV_i(z_i,\bar z_i)
\xi(z_i) \} -\p\big(\hV_i(z_i,\bar z_i)c(z_i)\xi(z_i)\big)
-\bar \p\big(\hV_i(z_i,\bar z_i)\bar c(z_i)\xi(z_i)\big)
\eqn\etwentysevenzb
$$
correspond to integrated vertex operators in the zero picture.
Finally,
$\hV_i(z_i,\bar z_i)c(z_i)\bar c(\bar z_i) X(z_i)$
correspond to unintegrated vertex operators in the zero picture.
Thus we see that the right hand side of eq.\etwentysevenza\ has precisely
the form expected from the analysis of the first quantized theory.


We now turn to the problem of determining $\Phi_N^{(r)}$ satisfying
eqs.\eseventeen\ and \eeighteen.
We shall discuss one particular construction, but one should keep in
mind that this construction is in no  way unique, and there are
(infinitely) many other choices possible.
In order to prescribe $\Phi_N^{(r)}$ in a way that avoids the divergences
associated with collision of picture changing
operators\WENDT,\foot{Removal
of such divergences in open string field theory has been discuss by
several authors\con\MATV\SAMUEL\THORN\noc} we shall find it
more
convenient to take the $N$-string diagram not just a regular polyhedra
with $N$ faces, but regular polyhedra with $N$ faces with tubes of a
certain fixed length $l_0$ attached to each of the faces; together with
diagrams corresponding to such regular polyhedra with $m_1$, $m_2$,
$\ldots$,
$m_{r+1} =N+2r-(m_1+\ldots +m_r)$ faces, joined by tubes of length
$l_1,\ldots l_r$ ($0\le l_i\le 2l_0$) and twist $\theta_1,\ldots \theta_r$
($0\le\theta_i<2\pi$).
(Such vertices have been used in ref.\ZWIEBACH\ for a different purpose.)
We shall denote by $\rn_e$ the component of $\rn$ for which the
corresponding string diagram is a regular $N$-hedron with tubes of length
$l_0$ attached to its faces.
$\rn_c$ will denote the component of $\rn$ for which the corresponding
string diagram is given by two or more such regular polyhedra connected by
tubes of length $\le 2l_0$.

We shall first discuss the construction of $\Phi_N^{(0)}$ for points
inside $\rn_e$.
In this case the picture changing operators are taken to be at the
mid-points of the edges of the polyhedron.
Since the number of edges ($3N-6$) of an $N$-hedron is larger than the
number of picture changing operators ($N-2$), there are several
possibilities.
We average over all configurations, with the weight factor for a given
configuration being proportional to $\prod_i f(s_i)$, where $s_i$ is the
length of the $i$th edge, the product over $i$ runs over all the $(N-2)$
edges containing the picture changing operators, and $f(s_i)$ is a smooth
function of $s_i$ satisfying the constraint:
$$
f(s_i)=0 ~~{\rm for}~s_i\le\eta;~~~ f(s_i)=1~~{\rm for}~s_i\ge 2\eta
\eqn\elione
$$
where $\eta$ is a small but fixed number.
This construction guarantees that the picture changing operators are
always inserted at the mid-points of the edges which have length $\ge
\eta$, and hence two picture changing operators never collide inside
$\rn_e$. \foot{ Note that we could also have, in principle, chosen the
picture changing operators at the vertices, imitating the corresponding
construction for open string field theory.
But in this case special care (like moving the picture changing operators
away from the vertices) is needed to ensure that two picture changing
operators do not collide, since for $N\ge 5$, $\rn_e$ contains
configurations where the number of well separated vertices (say, by a
distance $\ge\eta$) is less than $N-2$.
Such a situation does not occur if we insert the picture changing
operators at the midpoints of the edges, since a regular N-hedron contains
at least $N$ edges of finite length.
}
This completely specifies $\Phi_N^{(0)}$ inside $\rn_e$.

Let us now turn to a point inside $\rn_c$ which corresponds to two regular
polyhedra joined by a tube of length $l$.
Let $h$ be some fixed length $\le l_0$.
For $l\ge 2h$, we choose $\Phi_N^{(0)}$ to be simply the product of
$\Phi_N^{(0)}$ on the two polyhedra.
This automatically ensures that $\Phi_N^{(0)}$ satisfies the boundary
condition \eeighteen\ at the boundary $l=2l_0$ of $\rn$.
For $l\le h$, we choose the picture changing operators to be on the edges
of the two polyhedra in such a way that $\Phi_N^{(0)}$ is independent of
$l$ and is identical to its value at $l=0$, where it is given by the
previous construction of $\Phi_N^{(0)}$ for points inside $\rn_e$.
This does not specify $\Phi_N^{(0)}$ completely, since a picture changing
operator inserted on the right boundary of the tube or the left boundary
of the tube at the same angle generates the same configuration when the
tube length is collapsed to zero.
We remove this ambiguity by assigning equal weight factor to each of these
configurations.
Finally, in the region $h\le l\le 2h$, $\Phi_N^{(0)}$ is chosen to be a
linear combination of the expressions for $\Phi_N^{(0)}$ for $l\le h$ and
$l\ge 2h$ such that the resulting expression smoothly interpolates between
the values of $\Phi_N^{(0)}$ for $l\le h$ and $l\ge 2h$.

This construction can easily be generalized to points inside $\rn_c$
representing string diagrams where more than two polyhedra are joined
together by tubes of length $\le 2l_0$.
Ambiguities similar to the one discussed above arises in the region when
some of the tubes are of length $\le h$, since many different segments of
edges belonging to different polyhedra may correspond to the same segment
when all tubes of length $\le h$ is collapsed.
Hence if the polyhedron obtained after this collapse has an insertion of
picture changing operator on this segment, we need to decide how to
distribute it over the various segments before the collapse.
One possible consistent way is to distribute it only among the two
segments which are at the two extreme ends of the chain of tubes
connecting these different segments, with equal weight factor.

This finishes the discussion on the construction of $\Phi_N^{(0)}$.
Once $\Phi_N^{(0)}$ is given, $\Phi_N^{(r)}$ may be obtained by solving
the descent equations.
We shall not discuss the general case here, but as an example, give a
specific solution for $\Phi_4^{(r)}$.
In this case $\Phi_4^{(0)}$ may be expressed as
$$
\Phi_4^{(0)}=X(P)X(Q) +\sum_\alpha f^{(\alpha)}(\tau) X(P^\al) (X(Q^\al)
-X(R^\al))
\eqn\efourpointone
$$
where the sum over $\alpha$ runs over a finite set.
The corresponding solution for $\Phi_4^{(r)}$ is,
$$\eqalign{
\Phi_4^{(1)} =& (d\xi(P)X(Q)+X(P)d\xi(Q)) +\sum_\alpha df^\al X(P^\al)
(\xi(Q^\al) -\xi(R^\al))\cr
&+\sum_\alpha f^\al \Big(d\xi(P^\al)\big(X(Q^\al)-X(R^\al)\big) +X(P^\al)
\big(d\xi(Q^\al)
-d\xi(R^\al)\big)\Big)\cr
\Phi_4^{(2)} =& d\xi(P)\wedge d\xi(Q) -\sum_\alpha df^\al\wedge
d\xi(P^\al) (\xi(Q^\al) -\xi(R^\al))\cr
& +\sum_\alpha f^\al d\xi(P^\al)\wedge \big(d\xi(Q^\al)-d\xi(R^\al)\big)\cr
\Phi_4^{(r)} =& 0~~~~{\rm for}~r\ge 3
}
\eqn\efourpointtwo
$$
where $d\xi(P)=\p\xi(P) dP$ etc.


To summarize, in this paper we have discussed the construction of a gauge
invariant
field theory for the Neveu-Schwarz (NS) sector of the heterotic string
theory.
This construction can easily be generalized to include the NS-NS sector of
closed superstring field theory as well.
The Ramond sector, however, suffers from extra problems due to the
presence of zero modes of various operators, and cannot, at present, be
treated by the same method.
We hope to come back to this question in the future.

\refout

\end